\documentclass[aps,prb,twocolumn,groupedaddress,floatfix]{revtex4-1}

\usepackage{graphicx}
\usepackage{dcolumn}
\usepackage{bm}

\begin{document}

\title{Spin transition rates in  nanowire superlattices: Rashba spin-orbit coupling effects  }

\author{Sanjay Prabhakar,$^{1}$    Roderick Melnik,$^{1,2}$   and Luis L Bonilla$^{2,3}$}
\affiliation{
$^1$M\,$^2$NeT Laboratory, Wilfrid Laurier University, Waterloo, ON, N2L 3C5 Canada\\
$^2$Gregorio Millan Institute, Universidad Carlos III de Madrid, 28911, Leganes, Spain\\
$^3$School of Engineering and Applied Sciences,  Harvard University,  Cambridge, MA 02138 USA}

\date{May 21, 2013}

\begin{abstract}
We  investigate  the influence of Rashba spin-orbit coupling in a parabolic  nanowire modulated by  longitudinal periodic  potential. The modulation potential can be obtained from realistically grown supperlattices (SLs).  Our study shows that the Rashba spin-orbit interaction  induces the  level crossing point in  the parabolic nanowire SLs. We estimate large  anticrossing width (approximately 117 $\mu eV$) between singlet-triplet states. We  study the phonon and electromagnetic field mediated spin transition rates in the parabolic nanowire SLs. We report that the phonon mediated  spin transition rate is several order of  magnitude larger than the electromagnetic field mediated spin transition rate.
Based on the Feynman disentangling technique, we find the exact spin transition probability.  For the case  wave vector $k=0$, we report that the transition probability can be tuned in the form of resonance at fixed time interval. For the general case ($k\neq 0$), we solve the Riccati equation and  find  that the arbitrary values of $k$  induces the damping in the transition probability. At large value of  Rashba spin-orbit coupling coefficients for ($k\neq 0$), spin transition probability freezes.
\end{abstract}

\maketitle
\section{Introduction}
Low dimensional semiconductor nanostructures such as quantum dots, quantum wells and quantum wires  can be formed   in the plane of two-dimensional electron gas (2DEG) with the application of externally applied gate potentials,  have attracted significant interest for building robust spintronics logic devices and other applications.~\cite{prabhakar09,prabhakar10,erlingsson10,thorgilsson12,nefyodov11,prabhakar11,prabhakar12,glazov10,liang12}
Single electron spins  in these nanostructures can be manipulated by several parameters such as the gate controlled electric fields in the lateral direction and  externally applied magnetic fields. The Rashba and Dresselhaus spin-orbit couplings provide another  efficient way to control the single electron spins in these nanostructures.~\cite{prabhakar09,prabhakar11,takahashi10} The Rashba spin-orbit coupling arises due to structural inversion asymmetry in the crystal lattice along the growth direction.~\cite{bychkov84} The Dresselhaus spin-orbit coupling arises due to bulk inversion asymmetry  in the system.~\cite{dresselhaus55}

Accurate estimation of the spin transition  rate, mediated by phonons and electromagnetic fields, are of great interest for the design of optoelectronic  devices.~\cite{trif08,frantsuzov08,prabhakar12}
Long spin relaxation rates, approximately $0.85$ ms
in GaAs quantum dots and $20$ ms in InGaAs quantum dots,  have been measured by utilizing several different experimental techniques such as pulsed relaxation and optical orientation methods.~\cite{elzerman04,kroutvar04} In  these experiments, it is confirmed that the transition  rate is dominated  by  the spin-orbit coupling
with respect to the environment.~\cite{golovach04,khaetskii00,prabhakar10}
Because of the spin-orbit coupling, the electron spin qubits in the  nanowire quantum dots localized in a  transmission line resonator can be manipulated, stored and read out with the application of the gate controlled electric fields.~\cite{trif08,frey12}
In this paper, we  investigate the energy spectrum of the parabolic nanowire modulated by realistically grown SLs in the longitudinal direction.~\cite{schomburg98,voon03,bastard-book} We focus on the study of  the crossing of the energy spectrum  of the nanowire SLs accounting for  the Rashba spin-orbit coupling. The crossing point can be achieved with the accessible values of the strength of the Rashba spin-orbit coupling. We investigate the spin transition rate in parabolic nanowire SLs with the Rashba spin-orbit coupling under the influence of  electromagnetic field radiation,~\cite{pietilainen06,dugaev09} phonons~\cite{khaetskii00} and Dyakonov-Perel (DP)~\cite{kiselev00} mechanisms. We can write the momentum as a classical variable in the Dyakonov-Perel mechanism  under  the  Markovian process~\cite{marco05,kiselev00,schliemann03} and estimate the transition probability by utilizing the Feynman disentangling technique method.~\cite{feynman51,prabhakar10,popov07} The DP mechanism corresponds to  the spin splitting of the conduction band in zinc blende semiconductors at finite wave vectors which is  equivalent to the presence of an effective magnetic field that causes  the precession of an  electron spin.~\cite{marco05,kiselev00}

The paper is organized as follows: in section~\ref{theoretical-model}, we develop a theoretical model that allows us to find   the crossing of the  energy spectrum  of the  parabolic nanowire SLs with Rashba spin-orbit coupling. In this section, we also develop the theoretical model that allows us to find the spin  transition rate. In  section~\ref{results}, we plot the dispersion relation (see Figs.~\ref{fig1}) of the parabolic nanowire modulated by longitudinal periodic potential. In Fig.~\ref{fig2} and \ref{fig3}, we  plot the spin transition rate of the nanowire SLs with Rashba spin-orbit coupling  via   electromagnetic field radiation, phonons and Dyakonov-Perel mechanisms.   Finally, in section~\ref{conclusion}, we summarize our results.

\section{Theoretical Model}\label{theoretical-model}
\begin{figure}
\includegraphics[width=8cm,height=7cm]{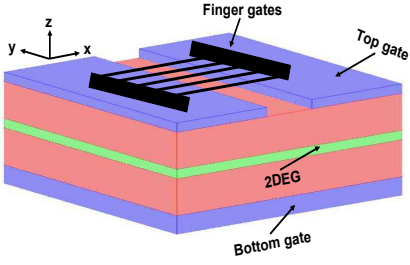}
\caption{\label{fig1} (Color online) Schematic diagram of the proposed experimental setup of the parabolic NWSLs periodically modulated by an external potentials via finger gates. The 2DEG can be realized in InAs quantum well by sandwitching between two GaAs barrier materials. The top gate induces the parabolic NWSLs while the bottom gate control the strength of the Rashba spin-orbit coupling.   }
\end{figure}

\begin{figure*}
\includegraphics[width=18.5cm,height=7.5cm]{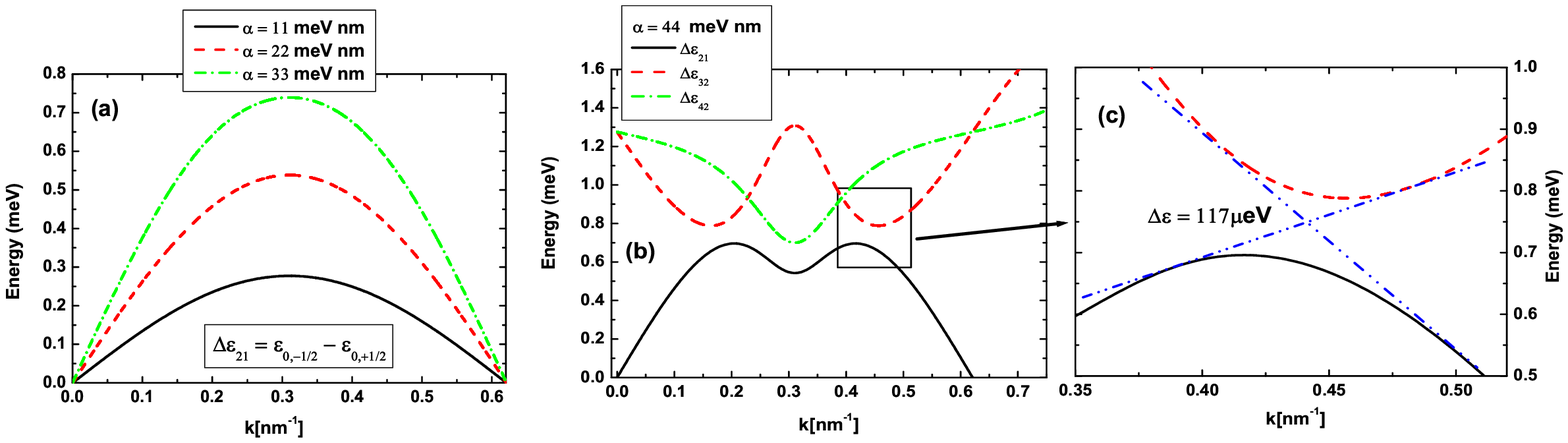}
\caption{\label{fig2} (Color online) Dispersion relation: energy diff.  vs $k$. (a) spin splitting energy increases and becomes maximum at $k=\pi/2l$. (b) It can be seen that the crossing between the states $|0,-1/2\rangle$ and  $|1,+1/2\rangle$  takes place approximately at $k=0.19/\mathrm{nm}$ and $k=0.43/\mathrm{nm}$. Here we find the anticrossing width is approximately $117~\mathrm{\mu eV}$. We chose the material constants for InAs material as $m=0.0239$, $\hbar\omega_0=1.3~\mathrm{meV}$, $\Delta_1=16~\mathrm{meV}$ and $l=5.06~\mathrm{nm}$.    }
\end{figure*}

\begin{figure*}
\includegraphics[width=16cm,height=7.5cm]{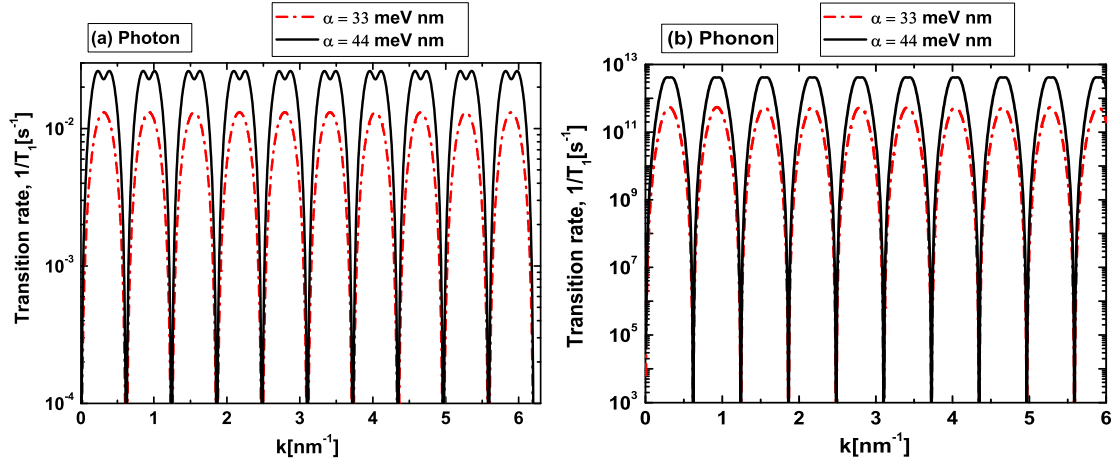}
\caption{\label{fig3} (Color online) (a) Electromagnetic field mediated spin transition rate vs $k$ in InAs nanowire SLs.  In fact, the transition rate depends on the amplitude of the external field, if it is classical. However, since it is taken quantized in Eq.~(\ref{Art}), the field amplitude does not influence Fig.~\ref{fig2}(a).    (b) Phonon mediated spin transition rate vs $k$. The dips in the transition rate (see solid lines)  can be seen due to level crossing. Here we chose $\hbar\omega_0=1.3~\mathrm{meV}$, $eh_{14}=0.54\times 10^{-5} ~\mathrm{erg/cm}$, $s_l=4.2\times 10^{5} ~\mathrm{cm/s}$, $s_t=2.35\times 10^{5}~\mathrm{cm/s}$, $\rho=5.6670 ~\mathrm{g/cm^3}$, $\epsilon_r=14.6$, $\Delta_1=16~\mathrm{meV}$ and $l=5.06~\mathrm{nm}$.  }
\end{figure*}

The total Hamiltonian of the quasi one dimensional parabolic  nanowire formed in the  plane of 2DEG  with  Rashba spin-orbit coupling in presence of longitudinal  modulation potential (see Fig.~\ref{fig1} for experimental set up) can be written as~\cite{kleinert05,marinescu10,thorgilsson12,lutchyn10,sherman12}
\begin{eqnarray}
H=H_{xy}+H_z+H_R,\label{total}\\
H_{xy}=\frac{p_y^2}{2m}+\frac{1}{2}m\omega_0^2y^2+\varepsilon_\nu(k),\label{Hxy}\\
H_z=p_z^2/2m+V(z),\label{Hz}\\
H_R=\frac{\alpha}{\hbar}\left(p_y\sigma_x-p_x\sigma_y\right),\label{HR}
\end{eqnarray}
where $H_z$ is the Hamiltonian of the electrons along z-direction and $H_R$ is the spin-orbit Hamiltonian associated to the Rashba coupling. The full descriptions of $H_z$ and $H_R$ will be discussed shortly.  In the above Hamiltonians $p_x$, $p_y$ and $p_z$ are the momentum operator, $m$ is the effective mass, $\alpha$ is the strength of the Rashba spin-orbit coupling and $\omega_0=\hbar/m\ell_0^2$ is the confinement frequency of the parabolic potential with $\ell_0$ being the oscillator strength.   $\sigma_i~ (i=x,y,z)$  are the Pauli spin matrices.  $\varepsilon_\nu(k)$ provides the periodic longitudinal modulation potential along x direction  in the form of:~\cite{thorgilsson12,marinescu12,bonilla11}
\begin{equation}
\varepsilon_\nu(k) =\frac{\Delta_1}{2}\left(1-\cos kl\right),
\label{varepsilon-k}
\end{equation}
where $\Delta_1$ is the first miniband width and $l$ is the superlattice period.~\cite{voon03,willatzen04} The second term in ~(\ref{total}) represents the Hamiltonian of the electron along  z-direction  where $V_z$ is the asymmetric triangular quantum well confining potential along z-direction. Usually, the asymmetric triangular quantum well potential can be found by solving the Schr$\mathrm{\ddot{o}}$dinger-Poisson equations self-consistently.~\cite{prabhakar09,stern84} The potential along z-direction can be chosen as  $V_z=eEz$ for $z\geq 0$ and $V_z=\infty$ for $z< 0$.~\cite{sousa03} The ground state wavefunction ($\Psi_{0z}(z)$) of $H_z$ can be written in the form of Airy function ($\mathrm{Ai}$)  as~\cite{prabhakar09,stern84,sousa03}
\begin{equation}
\Psi_{0z}(z)=1.4261 q^{1/2}\mathrm{ Ai}\left(qz+\varphi_1\right),\label{psi}
\end{equation}
where   $\varphi_1=-2.3381$ is the first zero of the Airy function and
\begin{equation}
q=\left[\frac{2meE}{\hbar^2}\right]^{1/3}.
\end{equation}

We will make use of an average momentum squared in the state~(\ref{psi}), $\langle p_z^2\rangle=0.78 \left(\hbar q\right)^2$,  and the average position  $\langle z\rangle =1.56/q$ to estimate   the thickness of the $2$DEG. The structural inversion asymmetry in $V_z$ leads to the Rashba spin-orbit coupling (see Eq.~\ref{HR}) and the Rashba coefficient $\alpha$ can be written as~\cite{silva97}
\begin{eqnarray}
\alpha=\frac{\gamma_R e\langle E \rangle}{\hbar},
\gamma_R=\frac{\hbar^2\Delta\left(2E_g+\Delta\right)}{2mE_g\left(E_g+\Delta\right)\left(3E_g+2\Delta\right)},~\label{gamma-R}
\end{eqnarray}
where $\Delta$ stands for the spin-orbit splitting in the valence band and $E_g$ is the band gap. For InAs material, we adopt the value $\gamma_R=110 {\AA}^2$.

Since,  $[p_x,H_{xy}]=0$, we  consider  $p_x$ is the good quantum number and the eigenvalue of $p_x$ can be written as~\cite{marinescu10,kleinert05,bonilla11}
\begin{equation}
p_x=\frac{m}{\hbar}\frac{d\varepsilon_v(k)}{dk}=\frac{ml}{2\hbar} \Delta_1 \sin kl.
\label{p-x}
\end{equation}

The Hamiltonians~(\ref{Hxy}) and (\ref{HR}) can be written in terms of annihilation and creation  operators as
\begin{eqnarray}
H_{xy}=\left(a^\dagger a+\frac{1}{2}\right)\hbar\omega_0+\varepsilon_\nu (k)\label{Hxy-1}\\
H_R=-\frac{\alpha m l}{2\hbar^2}\Delta_1 \sin kl~ \sigma_y+\frac{i\alpha}{\ell_0\sqrt 2}\left(a^\dagger - a\right)\sigma_x, \label{HR-1}
\end{eqnarray}
To find the energy spectrum of the above Hamiltonian, it is convenient to rotate the Hamiltonian $\tilde{H}_{xyR}=\exp{(-i\pi\sigma_x/4)}\left(H_{xy}+H_R\right)\exp{(i\pi\sigma_x/4)}$ so that the eigenvalue of  $p_x$ couples to $\sigma_z$. The new Hamiltonian can be written as $\tilde{H}_{xyR}=H_0+H_1$, where $H_0$ is the diagonal part and $H_1$ is the nondiagonal part.~\cite{thorgilsson12}
\begin{eqnarray}
H_0=\left(a^\dagger a+\frac{1}{2}\right)\hbar\omega_0-\frac{\Delta_1}{2}\frac{ l}{\xi} \sin kl~ \sigma_z+\varepsilon_\nu (k),~\label{H0}\\
H_1=\frac{i\alpha}{\ell_0\sqrt 8}\left(a^\dagger - a\right)\sigma_+ + H.c.,~\label{H1}
\end{eqnarray}
where $\xi=\hbar^2/(\alpha m)$ is the spin precession  length, $\sigma_{\pm}=\sigma_x\pm i \sigma_y$ and $H.c.$ stands the Hermitian conjugate. From Eq.~(\ref{H1}), it is clear that the non-diagonal part couples the state differed by one quantum number. In a case, where $H_0 \gg H_1$, we use non-diagonal Hamiltonian $H_1$ as a perturbation. Based on the second order perturbation theory, the energy spectrum of the nanowire  can be written as
\begin{eqnarray}
\varepsilon_{n,+1/2}&=&\left(n+\frac{1}{2}\right)\hbar\omega_0-\frac{\Delta_1}{2}\frac{ l}{\xi} \sin kl +\varepsilon_\nu(k)+\nonumber\\
&&\frac{\alpha^2\xi}{8\ell_0^2}\left[\frac{n}{\hbar\omega_0\xi-\Delta_1 l\sin kl}-\frac{n+1}{\hbar\omega_0\xi+ \Delta_1 l\sin kl}\right], ~~~~~\label{E}\\
\varepsilon_{n,-1/2}&=&\left(n+\frac{1}{2}\right)\hbar\omega_0+\frac{\Delta_1}{2}\frac{ l}{\xi} \sin kl +\varepsilon_\nu (k)+\nonumber\\
&&\frac{\alpha^2\xi}{8\ell_0^2}\left[\frac{n}{\hbar\omega_0\xi+\Delta_1 l\sin kl}-\frac{n+1}{\hbar\omega_0\xi-\Delta_1 l\sin kl}\right]. \label{E1}
\end{eqnarray}
We now turn to the calculation of spin-flip transition rate in parabolic nanowire modulated by periodic potential.
\begin{figure*}
\includegraphics[width=16cm,height=8.5cm]{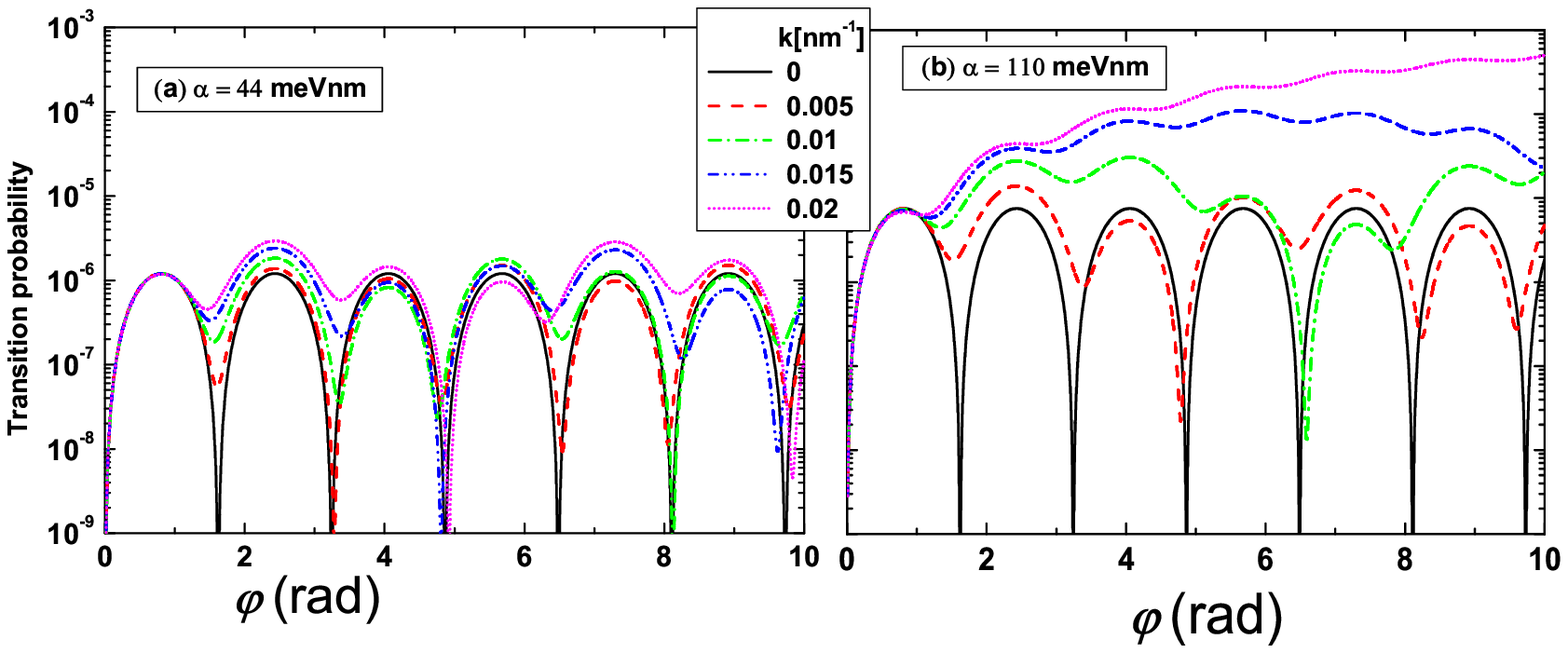}
\caption{\label{fig4} (Color online) Transition probability, $w_{1/2,-1/2}$ vs $\varphi$ in the InAs nanowire SLs. Here we chose $\Delta_1=72 ~\mathrm{meV}$, $l=53 ~\mathrm{nm}$, $\hbar\omega_0=1.3~\mathrm{meV}$,   $\xi_0=1.5a_0$ and $a_0=0.053~\mathrm{nm}$ is the atomic Bohr radius.   }
\end{figure*}

\subsection{Electromagnetic field mediated spin transition rate}
We apply the formalism of time dependent perturbation theory to the interactions of parabolic nanowire modulated by longitudinal periodic potential with the classical electromagnetic radiation field.~\cite{sakurai-book,bastard-book} The total Hamiltonian of the nanowire SLs with electromagnetic field radiation is given by~\cite{pietilainen06,avetisyan12}
\begin{eqnarray}
H=H_0+H_1+H_A,\label{H}\\
H_A=-\frac{e}{m}\mathbf{A}\cdot \mathbf{p}-\frac{\alpha e}{\hbar} A_y \sigma_x.\label{HA}
\end{eqnarray}
Eq.~(\ref{HA}) is treated  as an external perturbation dependent on the position $\mathbf{r}$ and time $t$.  The vector potential $\mathbf{A}$ ($\mathbf{r}$,t)  of the electromagnetic field radiation is written as
\begin{equation}
\mathbf{A}(\mathbf{r},t)=  \sqrt{\frac{\hbar}{2\epsilon_r \omega_{\mathbf{q}}V}} \hat{e}_{q\lambda} b_{q,\lambda} e^{i\left(\mathbf{q\cdot r} -\omega_q t\right)} + H.c.,
\label{Art}
\end{equation}
where $\omega_q=c|\mathbf{q}|$,  $b_{q,\lambda}$ annihilate photons with wave vector $\mathrm{q}$, $c$ is the velocity of light, $V$ is the volume of the nanowire,  $\epsilon_r$ is the dielectric constant of the nanowire. The polarization $\hat{e}_{q\lambda}$ with $\lambda=1,2$ are chosen as two perpendicular  induced photon modes  in the nanowire. The polarization directions of the induced photon are $\hat{e}_{q1}=\left( \sin\phi,  -\cos\phi, 0 \right)$ and  $\hat{e}_{q2}=\left(\cos\theta \cos\phi, \cos\theta \sin\phi, -\sin\theta \right)$ because we express $\mathbf{q}=q\left(\sin\theta \cos\phi, \sin\theta \sin\phi, \cos\theta \right)$. The above polarization vectors holds the relation as $\hat{e}_{q1}=\hat{e}_{q2}\times \mathbf{\hat{q}}$, $\hat{e}_{q2}= \mathbf{\hat{q}}\times \hat{e}_{q1}$  and $\mathbf{\hat{q}=\hat{e}_{q1}\times \hat{e}_{q2}}$. Based on the Fermi Golden Rule, the electromagnetic field mediated  spin transition rate ($i.e.,$ the transition probability per unit time) in the nanowire modulated by longitudinal periodic potential  is given by~\cite{sakurai-book}
\begin{eqnarray}
\frac{1}{T_1}&=&\frac{V}{\left(2\pi\right)^2\hbar}\nonumber\\
&&\int d^3\mathbf{q}\sum_{\lambda=1,2}\arrowvert M_{q,\lambda}\arrowvert^2\delta\left(\hbar\omega_\mathbf{q}-\varepsilon_{0,-1/2}+\varepsilon_{0,+1/2}\right),~~~~~
\label{1-T1}
\end{eqnarray}
where the matrix element $M_{q,\lambda}=\langle n,1/2|H_A|n,-1/2\rangle$ has been found perturbatively. The  spin transition rate   (i.e., $\left(n,-1/2\right)\rightarrow \left(n,+1/2\right)$) is given by
\begin{equation}
\frac{1}{T_1}=\frac{\alpha^2e^2\Delta_1 l \sin kl}{4\pi\hbar^4\xi\epsilon_r\epsilon_0c^3}\left[1-\frac{\alpha^2\xi^2}{4\ell_o^2\left\{ \left(\hbar\omega_0\xi\right)^2-\left(\Delta_1 l \sin kl\right)^2 \right\}}\right].
\label{1-T1-1}
\end{equation}

\subsection{Phonon mediated spin transition rate}
We now turn to the calculation of the phonon mediated
spin relaxation rate  in parabolic nanowire SLs. Following Ref.~\onlinecite{prabhakar12}, the interaction between electron
and piezo-phonon can be written as~\cite{khaetskii00,khaetskii01,woods02,olendski07}
\begin{equation}
u^{\mathbf{q}\alpha}_{ph}\left(\mathbf{r},t\right)=\sqrt{\frac{\hbar}{2\rho V \omega_{\mathbf{q}\alpha}}} e^{i\left(\mathbf{q\cdot r} -\omega_{q\alpha} t\right)}e A_{\mathbf{q}\alpha}b^{\dag}_{\mathbf{q}\alpha} + H.c.
\label{u}
\end{equation}
Here, $\rho$ is the crystal mass density, $V$ is the volume of the nanowire SLs, $b^{\dag}_{\mathbf{q}\alpha}$ creates an acoustic phonon with wave vector $\mathbf{q}$ and polarization $\hat{e}_\alpha$, where $\alpha=l,t_1,t_2$ are chosen as one longitudinal  and two transverse phonon modes. Also,  $A_{\mathbf{q}\alpha}=\hat{q}_i\hat{q}_k e\beta_{ijk} e^j_{\mathbf{q}\alpha}$ is the amplitude of the electric field created by phonon strain, where $\hat{\mathbf{q}}=\mathbf{q}/q$ and $e\beta_{ijk}=eh_{14}$ for $i\neq k, i\neq j, j\neq k$. The polarization directions of the induced phonon are $\hat{e}_l=\left(\sin\theta \cos\phi, \sin\theta \sin\phi, \cos\theta \right)$, $\hat{e}_{t_1}=\left( \sin\phi,  -\cos\phi, 0 \right)$ and $\hat{e}_{t_2}=\left(\cos\theta \cos\phi, \cos\theta \sin\phi, -\sin\theta \right)$. Based on the Fermi Golden Rule, the phonon induced spin transition rate in the nanowire SLs is given by~\cite{sousa03,khaetskii01}
\begin{eqnarray}
\frac{1}{T_1}&=&\frac{\left(eh_{14}\right)^2\alpha^2\left(\Delta_1 l \sin kl\right)^5}{70\pi \hbar^4\rho\xi \left[\left(\hbar\omega_0\xi\right)^2-\left(\Delta_1 l \sin kl\right)^2\right]^2}\left(\frac{1}{s^5_l}+\frac{4}{3s^5_t}\right)\nonumber\\
&&\left[1-\frac{\alpha^2\xi^2}{4\ell_o^2\left\{ \left(\hbar\omega_0\xi\right)^2-\left(\Delta_1 l \sin kl\right)^2 \right\}}\right]^3.
\label{1-T1-2}
\end{eqnarray}

\subsection{Spin-flip transition probability: Feynman disentangling technique}
We used the Feynman disentangling technique~\cite{feynman51,popov07,prabhakar10} to find the transition probability of electron spins of the Hamiltonian associated to  the Rashba spin-orbit coupling. Under the Dyakonov-Perel mechanism,~\cite{marco05,kiselev00} we consider  the momentum $\mathbf{p}(t)=m\dot{\mathbf{r}}$ as a classical variable whose dependence on time $t$ is generated by a Markovian process.~\cite{schliemann03}  In this case, the Rashba spin-orbit Hamiltonian $\tilde{H}_{R}=\exp{(-i\pi\sigma_x/4)}H_{R}\exp{(i\pi\sigma_x/4)}$ can be written as
\begin{equation}
\tilde{H}_{R}(t)=\frac{\xi_0}{\xi}\hbar\omega_0\cos\omega_0t \left(s_++s_-\right)-\frac{l}{\xi}\Delta_1 \sin kl ~s_z,\label{HR-2}
\end{equation}
where  $\xi=\hbar\omega_0\ell_0^2/\alpha$ is the spin-precession length,  $\xi_0$ is the spin-orbital radius and  $s_{\pm}=s_x\pm is_y$. The spin operators obey the $SU(2)$ algebra, $[s_+,s_-]=2s_z$ and $[s_z,s_{\pm}]=\pm s_{\pm}$. The spin evolution operator, $U(0,t)=T~\exp{\left(-i/\hbar\int dt \tilde{H}_{R}(t)\right)}$ can be exactly found by utilizing Feynman disentangling method~\cite{popov07,prabhakar10} $U(0,t)=\exp{\left(a(t)s_+\right)}\exp{\left(b(t)s_z\right)}\exp{\left(c(t)s_-\right)}$, where $a(t), b(t)$ and $c(t)$ are time dependent functions that can be found exactly. In the disentangled form, the evolution operator can be written as
\begin{widetext}
\begin{equation}
U(t)= e^{a(t)s_+}  T \exp\left\{{-\frac{i}{\hbar}\int^t_{t'=0}\left[\left(\frac{\xi_0}{\xi}\hbar\omega_0\cos\omega_0t-\chi\right)s'_+ -\frac{l}{\xi}\Delta_1 \sin kl ~s'_z+\frac{\xi_0}{\xi}\hbar\omega_0\cos\omega_0t ~s'_-\right]dt'}\right\},   \label{U2-1}
\end{equation}
\end{widetext}
where $T$ is the time ordering operator and
\begin{equation}
a(t)=-\frac{i}{\hbar}\int\chi(t')dt',\label{at-1}
\end{equation}
\begin{equation}
s'_\mu=\exp\left\{{\frac{i}{\hbar}s_+\int\chi(t')dt'}\right\}  s_\mu  \exp\left\{{-\frac{i}{\hbar}s_+\int\chi(t')dt'}\right\}.\label{s-mu}~~
\end{equation}
By differentiating Eq.~(\ref{s-mu}) with respect to `a' and utilizing the initial condition $s'_\mu(0)=s_\mu$, we find the relations $s'_+=s_+$, $s'_z=s_+a+s_z$,  $s'_-=s_--2s_za-a^2s_+$. By substituting these relations in Eq.~(\ref{U2-1}) and equating the coefficient of $s_+=0$, we find the differential equation in the form of
\begin{equation}
\frac{da}{dt}=-\frac{i}{\hbar}\left\{i\frac{\xi_0}{\xi} \hbar\omega_0\cos {\omega_0 t}\left(1-a^2\right)-a \frac{l}{\xi}\Delta_1 \sin kl \right\}.\label{da-dt}
\end{equation}
In an analogous way, we can disentangle $s_0$ and $s_-$ by differentiating Eq.~(\ref{s-mu}) with respect to `b' and `c' respectively. However, a single function $a(t)$ needed to find the transition probability.~\cite{popov07} The  differential Eq.~(\ref{da-dt}) can be solved exactly for the case $k=0$. Thus its solution can be written as
\begin{equation}
a(t)=\frac{\exp\left\{-2i\xi_0\sin \left(\omega_0 t\right)/\xi\right\}-1}{\exp\left\{-2i\xi_0\sin \left(\omega_0 t\right)/\xi\right\}+1}. \label{at}
\end{equation}
The transition probability between the opposite spin states can be written as
\begin{equation}
\omega_{+1/2, -1/2}=\frac{|a|^2}{1+|a|^2}=\sin^2\left(\frac{\xi_0}{\xi}\sin {\omega_0 t}\right).\label{w1}
\end{equation}
From Eq.~\ref{w1}, it is clear that the spin flip transition probability is enhanced with the Rashba spin-orbit coupling.
At large value of Rashba spin-orbit coupling constant,  the splitting of the peak value in the transition probability  can be achieved due to the fact that the periodicity of the propagating waves in the crystal lattice changes with $\alpha$. For the general case $k\neq 0$, we solve numerically the Riccati Eq.~(\ref{da-dt})  to find the transition probability.
In the disentangling operator scheme, as mentioned before, we write the momentum ($\mathbf{p}(t)=m\dot{\mathbf{r}}$) as a classical variable under  the Dyakonov-Perel mechanism,~\cite{marco05,kiselev00}  whose dependence on time $t$ is generated by a Markovian process.~\cite{schliemann03} Thus the Riccati Eq.~(\ref{da-dt}) can be treated as a nonrelativistic case. In such a situation, if a particle with a magnetic moment travels with a
relativistic speed in an electromagnetic field  whose orbital movement can be regarded as a classical then one might expect a change in the spin behavior (or in the particle polarization vector).~\cite{bargmann59} Thus from Eq.~(\ref{w1}), one might expect to flip the spin completely (resonance case)~\cite{rabi37} if $\sin\varphi=(n+1/2)\pi\xi/\xi_0$ where $\varphi=\omega_0t$ and $n=0,1,2,3\cdots$.

\section{Results and Discussions} \label{results}

In Fig.~\ref{fig2}(a), we have plotted energy vs $k$  of InAs parabolic nanowire SLs in presence of the Rashba spin-orbit coupling. We find the Kramer's type degeneracy point at $k=\pm n\pi/l$ with $n=0,1,2\cdots$. For the arbitrary values of $k$, the degeneracy can be lifted and we find the spin splitting energy. The splitted energy difference increases  with the Rashba spin-orbit coupling strength and becomes maximum at $k=\pi/2l$. Note that this maxima point is also maxima of $p_x$ (see Eq.~\ref{p-x}). In Fig.~\ref{fig2} (b), we investigate
the level crossing point associated with  the spin states $|n,-1/2\rangle$ and $|n+1,1/2\rangle$. The crossing  takes place at $k\approx 0.19/\mathrm{nm}$ and $k \approx 0.43/\mathrm{nm}$. Here  the  width of the anticrossing is approximately estimated as $117 \mathrm{\mu eV}$.

We now turn to the key results of the paper: the spin transition rate via electromagnetic field radiation, phonons and Dyakonov-Perel mechanism.~\cite{dugaev09,khaetskii00,marco05,kiselev00}

In Fig.~\ref{fig3}, we plot the spin transition rate vs $k$ between the spin states $|0,-1/2\rangle$ and $|0,+1/2\rangle$ under the influence of electromagnetic field radiation and phonons. It can be seen that the phonon mediated spin transition rate is several order of magnitude larger than the value mediated by electromagnetic field radiation because the electromagnetic field mediated transition rate vanishes like $\sin kl$ whereas, the phonon mediated spin transition rate vanishes like $(\sin kl)^5$ (see Eqs.~\ref{1-T1-1} and~\ref{1-T1-2}). The dips in the  spin transition rate (see Fig.~\ref{fig2}, solid lines) can be seen due to level crossing.  The dips in the spin transition rate can be tuned with the application of the Rashba spin-orbit coupling.  The cusp like structure in the spin flip rate was reported by the authors in Refs.~\onlinecite{bulaev05,bulaev05a} for quantum dots system. For the case of parabolic nanowires  modulated by longitudinal periodic potential, the dips found in the spin transition rate  is new.

Finally, in Fig.~\ref{fig4}, we plotted the transition probability vs time.  In Fig.~\ref{fig4}(a) and (b), we show that at $k=0$ (see solid lines), the spin transition probability   can be tuned with fixed time interval in the form of resonance.
From Eq.~\ref{w1}, we can write the theoretical condition for finding zero transition probability for the case ($k=0$, solid lines (black) in Fig.~\ref{fig4}) as
\begin{equation}
\sin\varphi=n\pi\frac{\xi}{\xi_0},\label{sin-phi}
\end{equation}
where  $n=0,1,2,3\cdots$. The condition~(\ref{sin-phi}) is satisfied by solid lines (black) in Fig.~\ref{fig4} and thus we find the zero transition probability at fixed interval of $\varphi$.
Also, it can  be seen that the spin flip transition probability can be enhanced with the Rashba spin-orbit coupling.
Next, we  study the influence of the inclusion of the wavenumber ($k$). We find that there is a  superposition  effect between $p_x$ and $p_y$ in the evolution of the spin dynamics (see Eqs.~\ref{U2-1} and \ref{da-dt})  which induces the damping effect or spin echo  in the transition probability with respect to time.
For simplicity in Eq.~(\ref{da-dt}), we consider
\begin{equation}
\sin k \ell=\frac{\xi_0 \hbar\omega_0}{\Delta_1\ell} \cos\varphi.\label{sin-k}
\end{equation}
Here $\cos\varphi$ oscillates between -1 to +1 and for fixed value of $k$, the above condition~(\ref{sin-k}) is satisfied several times with the variation of $\varphi$. Thus,  one can find the zero spin transition probability at $\varphi=n\pi$. Whenever condition~(\ref{sin-k}) is violated (i.e., $\sin k \ell\neq \xi_0 \hbar\omega_0\cos\varphi/(\Delta_1\ell) $), then we find a superposition or spin echo (see Fig.~\ref{fig4} for the case $k\neq 0$).
At large value of the Rashba coefficient $\alpha$, the third term of Eq.~(\ref{da-dt}) dominates over the first and second term and we find the spin freezing  in the transition probability (see dashed line in Fig.~\ref{fig4}(b)). Recently, similar type of results have been shown in Ref.~\onlinecite{sherman12}.

\section{conclusion} \label{conclusion}
In Figs.~\ref{fig2}, we have demonstrated  that the level crossing in parabolic nanowire SLs  can be achieved with the accessible values of the strength of the Rashba spin-orbit coupling. The crossing point can be tuned to the lower values of $k$ and vice versa with the application of the Rashba spin-orbit coupling.
In Fig.~\ref{fig3}, we have shown that the phonon mediated spin transition rate is several order of magnitude larger than the electromagnetic field mediated spin transition rate. The dips in the  spin transition rate can be found due to level crossing.
Based on the Feynman disentangling technique method, in Fig.~\ref{fig4}, we have shown that the transition probability can be tuned in the form of resonance at  $k=0$. For the arbitrary values of $k$, we have shown that there is a superposition effect which induces the damping in the transition probability. At sufficiently large values of the Rashba spin-orbit coupling coefficients, the spin transition probability freezes for the arbitrary values of $k$. It means that one can not find the spin-flip transition rate at large value of $k$. In other words, manipulation (injection and detection) of  spin qubits far away from the gamma point in quantum wires is not the ideal candidate for the purpose of building a solid state quantum computer. It might be possible that our theoretically  investigated  spin-flip transition rate in the NWSLs can be experimentally measured   with current state of the art  technology (see Ref.~\onlinecite{liang12} for experimental set up).

\begin{acknowledgments}
This work has been supported by NSERC and CRC programs (Canada) and by FIS2011-28838-C02-01 (Spain).
\end{acknowledgments}

\end{document}